# Automating, Operationalizing and Productizing Journalistic Article Analysis

Eric Kingery
Public Good Software
Chicago, IL, USA
ekingery@publicgood.com

Michael S. Manley
Public Good Software
Chicago, IL, USA
msm@publicgood.com

Daniel Ratner
Public Good Software
Chicago, IL, USA
dan@publicgood.com

## ABSTRACT

Public Good Software's products match journalistic articles and other narrative content to relevant charitable causes and nonprofit organizations so that readers can take action on the issues raised by the articles' publishers. Previously an expensive and labor-intensive process, application of machine learning and other automated textual analyses now allow us to scale this matching process to the volume of content produced daily by multiple large national media outlets. This paper describes the development of a layered system of tactics working across a general news model that minimizes the need for human curation while maintaining the particular focus of concern for each individual publication. We present a number of general strategies for categorizing heterogenous texts, and suggest editorial and operational tactics for publishers to make their publications and individual content items more efficiently analyzed by automated systems.

## 1. Introduction: The Basic Problems

In 2013, Public Good Software (PGS) introduced a product designed to direct readers concerned with the causes they read about in journalism and other media with the nonprofit organizations working on those causes. For example, an article about a shooting in Chicago would link to local anti-violence organizations so a reader could donate, volunteer, or just spread the word. This product, called "Public Good for Media" (PGM) initially debuted as a Javascript-powered "Take Action Button" that publications could insert into the HTML source of any page (see Figure 1). This first iteration of the product required explicit configuration of each insertion by a publication's editorial staff with the assistance of PGS experts in curating the PGS marketplace's collection of organizations and causes. Explicit configuration quickly proved untenable for both the publications and PGS in terms of staff time and expense, publishing workflow modifications and lack of growth in traffic driven through the PGM program.

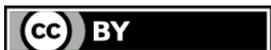

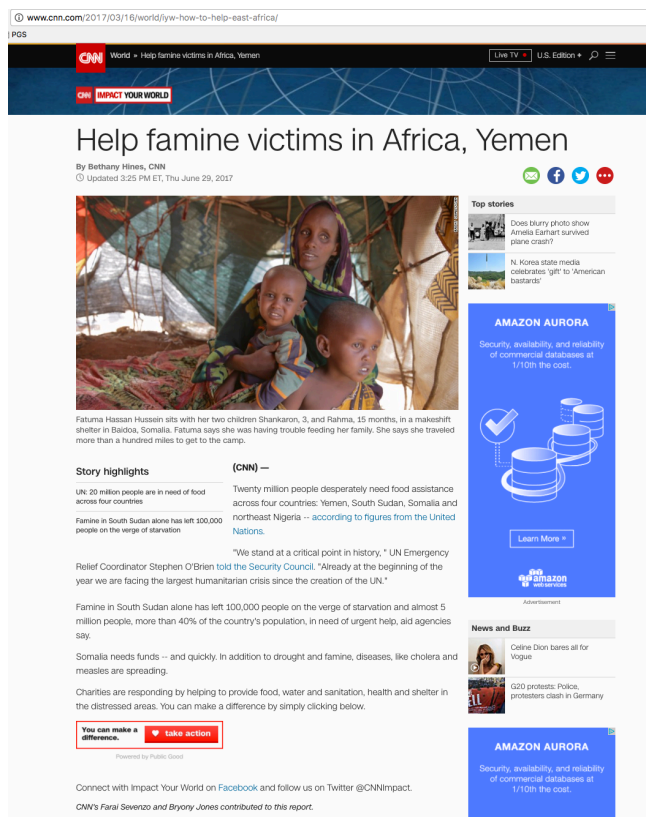

**Figure 1:** The Take Action Button embedded in an article

To address this, PGS created tools to automate the process so that the PGM product could be inserted into a site-wide template and relieve the media outlets' personnel of the burden of remembering how and when to include the PGM program as part of the publication process. Publishers also requested that PGS provide methods of automatically assigning the button's destination based on the content of each individual story, as well as the suppression of the button's display when there was no sensible destination, or when a story was not in any way cause-related (entertainment and sports, for example). Finally, while directing readers to nonprofits with a national area of effect was often sufficient, many publications either had more specifically located audiences, or wanted to highlight organizations working in the particular geographic area mentioned in the article.

These requests surfaced several questions and constraints for PGS to address in order to fully automate and operationalize the PGM product.

**Question: How can PGS automatically match an article to its underlying cause?** Causes may be in support of specific outcomes (for example, increased pet adoption) or against particular conditions (gun-related violence). In addition, PGS recognizes that causes often have a geographic area of effect (food security in the greater Chicago area). PGS has assembled a taxonomy of common causes, and then uses clues in the content of an article or the metadata about that article, to categorize articles in that taxonomy within some threshold of confidence. PGS must also identify the cases where it cannot provide a cause relevant to the content of an article, or when it must suppress any causes relevant to the article in accordance with the publisher's editorial stance. Given PGM's initial 37 categories of articles, random matching would yield only 2.7% accuracy, so any automated system would need to provide a much higher (20 times as high, minimally) rate of successful PGM insertions in order to be considered viable. In the absence of a standardized, machine-readable indicator of relevant causes, we require multiple strategies for categorizing these texts.

**Question: How can PGS match causes to the nonprofit organizations working on those causes?** Given a corpus of basic data about US nonprofit organizations provided by the IRS and supplemented by third-party data aggregation services and occasionally volunteered by the organizations themselves, PGS must identify the causes a given organization works on, or identify organizations whose work is relevant to given cause. There are over 1.5 million 501(c)(3) nonprofit organizations registered in the United States. Many of those organizations sponsor other cause-related projects, funds or community groups that PGM could surface as relevant to an article. Still more organizations do not concern themselves with cause-based activities or do not interact with the general public, and so must be excluded from consideration in the PGM program. Many organizations have separate geographic areas of effect from their administrative location. And the mission for many organizations is fluid and reactive to the needs of their areas of effect. There is no simple, deterministic method for mapping nonprofit organizations to the taxonomy of causes.

**Constraint: For the PGM program to grow to accommodate the volume of editorial content generated by multiple high-volume publishers on a daily basis, PGS must minimize human curation of answers to the above questions.** Some publishers generate dozens of pieces of content daily, and often syndicate that content to multiple partners. Other publishers apply the PGM program to their entire archive of content, adding to the sustained volume of matching requests. With the onboarding of only a handful of high-volume publishers, PGS could face a backlog of hundreds of articles per day that could potentially require human intervention to properly configure the PGM product on those articles. The labor cost involved with this stream of work would quickly become prohibitive to the program's growth.

The remainder of this paper discusses the approaches Public Good Software has taken to solving the problems of automating and operationalizing the Public Good for Media program so that it scales to support multiple high-volume journalism publishers while minimizing human curation costs.

## 2. The Initial Approach

When PGM initially launched, PGS staff worked with the staff of the publishers to determine which organizations or causes the system surfaced when a reader clicked the Take Action Button attached to an article. Once the IDs of the PGS marketplace entities were chosen, a web-based tool provided instructions and copy-and-paste code for the publisher's operations staff. A sample of that code is shown in Figure 2.

```
<script async type="text/javascript"
src="https://assets.pgs.io/pgm/v1/dpg.js">
</script>
<div class="pgs-dpg-btn" data-pgs-partner-
id="ucan" data-pgs-target-
type="takeaction/org" data-pgs-target-
id="ucan"></div>
```

**Figure 2: Sample code for the hand-curated PGM Take Action Button**

While the hand-curated insertion of PGM program code benefitted from nearly 100% accuracy in result relevance, it also maximized the amount of labor for both PGS and the publisher. This labor-intensive process remains an option for publishers who insert the PGM program into special features, or for publishers who only need to place the widgets in small volumes of content. Hand curation is operationally infeasible for most high-volume publishers.

To accommodate larger volumes of content, PGS modified its widget Javascript to send the URL of the article into which it was inserted to an "advice" service. That service performs multiple analyses on the text of the article, then uses the products of those analyses to match organization and cause entities in the PGS marketplace, or to decide to suppress display of the PGM widget entirely.

Our first analysis method consulted an early version of the service that would eventually become IBM's Watson Natural Language Understanding service [1] to isolate article text from the given URL's non-article content and to extract entities, geographic references, concepts and other semantic details from the isolated text. These extracted details acted as search terms against PGS's index of organization and cause entities, which included geographic terms, concepts and other data attached to the entities as "tags." This matching happened through an ElasticSearch index, which was also used to match the article text against "percolator" queries established for each organization name to catch direct mentions of organizations in the text.

This naive attempt at matching yielded an order of magnitude above random accuracy. However, this success rate of 25% alongside a non-trivial rate of false positive matches proved insufficient. Consequently, the human labor was shifted from pre-publication curation to post-publication quality assurance processes. Analysis of an article usually completed within a few seconds of it's first view. Subsequent views of the article would either hide or display the PGM widget configured with a set of actions for relevant organizations. This process triggered a notice of the analysis results to PGS staff. In the many results that were improperly matched, PGS staff used an internal QA tool to establish a "manual override" for that particular article which either suppressed the display of the widget permanently, or configured the action for that URL to more relevant results.

This first round of analyses raised several new issues:

- Articles often changed, sometimes dramatically, during the news cycle. As more and different information appeared in article text, the articles would need to be re-analyzed periodically. This often kicked off multiple rounds of manual intervention.

- Publishers often syndicated articles across different platforms, which meant that the same article text would be analyzed multiple times and trigger multiple manual interventions.

- Similarly, the same article could often appear at multiple URLs in the same publication, depending on the configuration of the publisher's content management system. Unfortunately, many content management systems unreliably or improperly implement the long-established mechanism for identifying duplicated content on the web, HTML's "canonical link" URL reference.

- Publishers often determined that some causes were outside the boundaries of their publication's concerns. They would express these preferences as either a list of "blacklisted tags" or "safelisted tags" to be used in deciding to display the PGM widgets. For the blacklist, the system needed to suppress the widget on a match. For the safelist, the system showed the widget only on a match.

Heavy application of blocklisted terms led to a successful match between articles and marketplace entities for between 30% and 40% of articles, but an accurate match only 25% of the time. In all other cases, widget display was suppressed unless manually overridden. PGS staff manually overrode more than half of all articles that passed through the system. At this point, we began looking for new analysis methods.

## 3. New Approach: Content Fingerprinting

A large component of the cost of human analysis of articles comes from the repeated analysis of the same piece of content appearing in different locations. When a piece of content appears in a single publication under multiple URLs, the PGM program first attempts to identify the article through a "canonical link" element found in the head of the HTML at the given URL. [2]

Unfortunately, many content management systems, both commercial and open source, appear to provide a poor canonicalization mechanism as implemented or configured in the production of many publications. Canonical links are frequently absent entirely, or may always be set to the URL used by the user-agent including some or all query string parameters. When articles are distributed via RSS/ATOM, Facebook Instant Articles, Accelerated Mobile Pages or other syndication forms, the absence of reliable canonical link URLs in any of these documents result in redundant analyses of the same content, and frequent multiple calls for human intervention on the same article.

For the articles missing canonical link URLs, PGS adopted an implementation of Moses Charikar's "SimHash" technique to identify duplicate content. [3] Before any article is submitted to automated analysis, the content of the article is extracted using the Newspaper library for Python [4] and a SimHash is computed. If the article's hash value is within a specified Hamming distance of a previously stored article, we can assume the articles are the same and we assign a previously calculated set of matching advice.

## 4. New Approach: Business Rules Engine

We approached the problems of properly classifying articles from multiple directions simultaneously. While investigating a number of machine learning techniques for classification, we sought to narrow the pool of articles requiring classification by applying deterministic analyses to the articles beforehand. While machine learning will always involve a margin of error, articles classified deterministically yield nearly 100% accuracy.

We continue to process the individual articles with the Newspaper library's textual analysis and natural language processing tools, which are similar to those provided by the Alchemy services. These analyses include entity extraction, geoparsing of location references, sentiment analysis and structural parsing of the article HTML source. The results from these analyses are appended to the metadata stored with each article.

We also analyze the state of the user agent, attempting to identify if the reader is currently logged in as a known user of PublicGood.com, where the user may be located (via whatever geolocation APIs may be available in the browser, or by IP address to location mapping services), and what previous actions the user may have taken during the current browsing session. The PGM widget Javascript collects this data as it is available and sends it along with the article URL to the advice service.

We generalized the blacklist and safelist solution into a form of advice configured by inspection of article metadata and markup. The individual inspections are codified in the form of publisher-specific productions for a forward-chaining business rules engine. These rules examine conditions present in the metadata and markup and also examine the current state of the advice generated by previous rules . This results in a set of marketplace entity references (causes and geo-specific organizations) and a widget show/hide flag. The widget Javascript consumes this advice to display the appropriate marketplace actions for the article.

Some rules are shared across multiple publishers, codifying common outcomes, such as suppressing widget display on articles shorter than a minimum word count. Other rules are publisher-specific, created through up-front manual inspection of the structure of a publication during a publisher's initial implementation of the PGM program. These inspections identify patterns in URL structure, article structure, and other clues that could be taken from publisher-provided metadata surrounding an article rather than the content of an article itself. Examples of these rules include identifying sections of a publication through URL patterns, or identifying topics and mappings to causes through a publication's own tagging and categorization strategy.

All of these rules take advantage of the value proposition of the business rules engine class of software. This allows non-technical personnel trained in publication analysis and rule writing to initially provision per-publisher rules and edit them according to business demands and publication changes.The costs of this human analysis can be limited in time and effort and made reactive to publisher changes through monitoring of pattern changes as well as structured communication with the publisher's staff. Matching patterns in metadata about the articles solves the entity matching problem for large portions of a publisher's content catalog for a much lower cost and faster implementation than would analyses of individual articles. Writing rules to attach human-curated results to specifically-identified articles also lessens the operational load on publisher staff, as they can implement the same PGM program code across all pages with the expectation that the rules established by PGS will properly assign widget actions.

## 5. Taxonomy & Metadata

Categorization of organizations, causes and articles about those causes requires a multi-layered set of taxonomies. Accurate matching becomes a problem of identifying intersections in the sets of taxonomy labels attached to those items. Finding those intersections is a fairly standard problem in data processing, which we solve through the application of document indexing and querying processes. In particular, PGS uses the ElasticSearch software package applied to JSON-serialized representations of its marketplace entities. These documents include not only all of the user-visible information about organizations, causes, campaigns and actions, but also multiple collections of supplemental data that serve to provide a richer view of the entities to search algorithms as well as machine learning processes.

Initially, PGS attempted to categorize nonprofit organizations in its marketplace using the National Taxonomy of Exempt Entities (NTEE) system maintained by the IRS and the National Center for Charitable Statistics. [5] While this taxonomy is useful for identifying large classes of organizations that should be excluded from matching (for example, most organizations in NTEE Category IX, Mutual/Membership Benefit), our experience has been that the 10 broad groups and 26 major groups described by the NTEE codes reported in nonprofit IRS filings are of less use when trying to determine more precisely what those organizations actually do and what causes they work on. In fact, NTEE classifications for an organization are often at odds with what the persons working for the organization consider as the organization's primary concern. This is particularly an issue for mature organizations whose mission has specialized or pivoted since their original filing with the IRS. So, while PGS still maintains NTEE classification as one facet of organizational data, that data is insufficient for matching purposes.

Another set of metadata very relevant to article-to-cause-to-organization matching is the set of geographic data for an organization. While an article may be parsed to extract geographic references, and the particular reader of an article may be geolocated through their device, their network address, or a location stored with PGS, geolocating an organization or a campaign is slightly more complex. In addition to the administrative location of an organization, we must also consider the area of effect for an organization's efforts. For example, the nonprofit Spark Ventures is headquartered in Chicago and does quite a bit of its fundraising there, but its primary areas of effect are in Nicaragua and Zambia. Other nonprofits may have more tightly clustered but still geographically disparate areas of effect. Still others have a global area of effect, with campaigns in specific locations as necessary. The metadata surrounding an organization must include all of these different locations and queries must specify which type of location is relevant. Are we looking for organizations based in Chicago, or are we looking for organizations helping people in Chicago?

Finally, there is the problem of categorizing articles. While entity extraction can occasionally identify specific nonprofit organizations (or their principals, through which the organization may be inferred), very often specific nonprofits are not the subject or even a tangentially mentioned detail in a cause-related story. This means we cannot rely on association between extracted entity and cause for categorizing a given article. For articles that cannot be categorized by their metadata and the business rules established for their surrounding publications, PGS must develop a taxonomy with which machines may be trained to classify articles. Here, we have a problem of balance in scope.

Categorizing free-form textual assets can quickly become a case of modeling the entire world. Take, for example, the category hierarchy for IBM's Natural Language Understanding product, which PGS used in its AlchemyLanguage product form [6]. This scheme provides almost 3,000 potential categorizations for an article. This is primarily useful for eliminating articles from consideration as cause-related at all. For example, practically all articles categorized as "art and entertainment : celebrity fan and gossip" can be excluded from consideration. Unfortunately, there are many more categories in this taxonomy to be ignored than to be useful, as very few of the categories are relevant to nonprofit organizations, causes and campaigns.

PGS uses supplemental data providers to attach additional taxonomies to marketplace entities (eg. United Nations Sustainable Development Goals [7] and media partner articles (e.g. AP News Taxonomy [8]. Machine learning classification models can then be trained and tested using the context-specific classes these taxonomies provide. Taxonomy development continues to progress, with several candidate taxonomies under development following consultation with several key publishers and a survey of the 200,000 articles already analyzed in the Public Good for Media program.

## 6. New approach: Machine Learning Models

Over half of the articles encountered in the current PGM corpus do not have enough information for the Business Rules Engine to deterministically match content to a cause. PGS originally relied on a manual QA process as described previously in Section 2. Given this manual intervention is a constraint on scaling the PGM program, machine learning was employed to assist in making match determinations in real time. With the help of advisor Q McCallum, we performed a broad survey of classification algorithms and methods and continue to explore new machine learning techniques and learn from the results we observe in production and development environments.

The initial approach employed a Naive Bayes classifier on a subset of violence-related news articles. This model performed exceptionally well, matching at over 80% accuracy for five similar yet distinct categories (Crime Prevention, Stop Gun Violence, Sexual Abuse Prevention, Child Abuse Prevention, Domestic Violence Prevention). However, applying the same technique to classify articles into a similar number of broader categories (violence-related, natural-disasters, health, etc.) resulted in an accuracy drop of 20% or more. After manually analyzing the training data, we concluded broader categories are prone to increased labeling ambiguity.

As we expanded the breadth of the classification model, we found that placing all articles unconcerned with causes into a single, broad category was not generalizable. Any positive results measured resulted from overfitting and/or a failure to properly shuffle the training data. In hindsight we identified multiple experiments that unknowingly used our training data labels to oversimplify the problem space.

With the scope of the ML solution expanded to more accurately match the problem space, a general-purpose Naive Bayes classification model was developed. This model encompasses the full 37-class news taxonomy referred to previously in Taxonomy & Metadata. It was trained and tested on roughly 6,000 manually labeled news articles selected from a small set of our publishers. Production results indicate an accuracy of over 50%.

We also surveyed and tested alternate algorithms (e.g. KNN, SVM, Neural Networks) along with multi-label and multi-model approaches. Some of these showed promise and improved on Naive Bayes results by 10-15%. Given the consistency of the results across various algorithmic approaches, we believe the influence of ambiguous and/or inconsistent training data is currently the largest limiting factor to increasing accuracy.

While we expect the machine learning models to improve as we implement new algorithms and continually train them with better-classified data and more robust taxonomies, we are currently considering a ceiling of 80% to 85% accuracy in cause matching through these methods. This underscores the importance of the Business Rules Engine and richer metadata around the marketplace entities. We expect to reach a point of diminishing returns in efforts to increase the ML accuracy, at which point we will be reliant on the Business Rules Engine to cut manual QA efforts, freeing our staff to perform more valuable work.

## 7. Conclusion

Over the past four years, Public Good Software has made significant strides in scaling the Public Good for Media program by automating the PGM cause matching product and expanding the analysis techniques used by that product to include both older, established software patterns (business rules engines) as well as newer computational methods (simhashing, machine learning algorithms). Given the expensive human effort required to properly match individual articles to their underlying causes and those causes to relevant actions with nonprofit organizations, PGS has found it more effective to focus human curation efforts in up-front analysis of general publication structure and in training machine learning models. By reducing the amount of human curation to only those instances where special attention is warranted (small content volume with high visibility, such as sponsored content packages) or when the automated systems cannot match with the desired confidence, PGS minimizes the cost of operating the PGM product at scale serving many high-volume publishers simultaneously. There are still a number of improvements to be made to the automated systems. PGS continues to test and train multiple machine learning models with a growing corpus of properly labeled and matched articles. We continue to refine the general nonprofit cause taxonomy and layer in additional sources of metadata about nonprofit organizations. As an example of investigating where these two areas intersect, we plan to run the categorization models against the web presences (sites, blogs, social media) of the nonprofits themselves.

We also intend to investigate further methods of deferring human curation, at least as it would apply to either PGS or publisher personnel. We believe there exists an opportunity to treat the classification of articles as a game that provides social and financial benefits to publication readers, to nonprofits, and to PGS and its media partners.

Finally, we intend to continue to work with our publisher partners and with publishers in general to find the best practices for structuring their publications in ways that make their articles more reliably analyzed and matched to relevant causes, organizations and actions. Suggestions for publishers include:

- Cataloging and normalizing URL patterns in their publications, and providing references to that catalog in a robots.txt, sitemap or other metadata resource.

- Assigning persistent content identifiers to their articles that follow a piece of content through multiple contexts and syndication formats. These could be reliable canonical link URLs, exposure of a publisher-specific content identifier, or the use of third-party handles on content items, such a Digital Object Identifiers.

- "Tagging" and categorizing content reliably with extracted entity names, publisher-specific causes or concerns, or with PGS or other third-party taxonomy terms.

- Marking up content with standard microformats, such as the Associated Press News markup format.

- Providing full-content syndication feeds, even if those feeds are not made public and are restricted to specific consumers, such as the PGM system.

- Paying close attention to the best practices for web publications as suggested by the prominent search engines. Efforts made to optimize search performance will also improve analysis of articles by systems like PGM.

## Notes & References

1: https://www.ibm.com/watson/services/natural-language-understanding/

2: https://en.wikipedia.org/wiki/Canonical_link_element

3: https://en.wikipedia.org/wiki/SimHash

4: http://newspaper.readthedocs.io/en/latest/

5: http://nccs.urban.org/classification/national-taxonomy-exempt-entities

6: https://www.ibm.com/watson/developercloud/doc/natural-language-understanding/categories.html

7: http://www.un.org/sustainabledevelopment/sustainable-development-goals/

8: https://developer.ap.org/ap-metadata-services#apNewsTaxonomy